\documentclass[10pt]{article}

\usepackage{xcolor}

\usepackage{amsmath}
\usepackage{amssymb}
\usepackage{graphicx,scalerel}
\usepackage{extsizes}
\usepackage{array}
\usepackage{url}
\usepackage{natbib}
\usepackage{bm}
\usepackage{caption}
\usepackage{float}
\usepackage{enumitem} 
\usepackage{subcaption}
\usepackage{booktabs}
\usepackage{caption}
\usepackage{lipsum}
\usepackage{framed}   
\usepackage{multirow} 
\usepackage{makecell}
\usepackage{bm}
\usepackage{tabularx}
\usepackage{array}
\usepackage{ragged2e}
\usepackage{array}
\usepackage{tikz}
\usetikzlibrary{positioning, shapes.misc, calc}
\usepackage{float}
\usepackage{caption}
\usepackage{amsmath}
\usepackage{threeparttablex}
\usepackage{pdflscape}
\usepackage{longtable}
\usepackage{array}
\usepackage{ragged2e}

\newcolumntype{Y}{>{\RaggedRight\arraybackslash}X}
\newcolumntype{L}[1]{>{\RaggedRight\arraybackslash}p{#1}}

\usepackage[a4paper, margin=2cm]{geometry}

\begin{document}

%\begin{titlepage}
\begin{center}
{\large\textbf{A framework for classifying and visualising experimental designs when subjects are measured repeatedly}}\\[1ex]

Damianos Michaelides$^{1}$,
Simon T. Bate$^{2}$\\[2ex]

$^{1}$ Department of Biostatistics, The Cyprus Institute of Neurology and Genetics, 1683 Nicosia, Cyprus\\
$^{2}$ CMC Statistical Sciences, GlaxoSmithKline, Stevenage, UK
\end{center}			
		%\vspace{3.8\baselineskip}
				
%\begin{abstract}

\normalsize

When running experiments that involve humans or animals, for example in clinical or pre-clinical research, it is often the case that multiple measurements are taken on the experimental subjects. This may involve measuring subjects repeatedly over time to track any trends, administering a sequence of treatments to compare their effects within-subject, or taking multiple technical replicate samples of each subject to obtain a more reliable average response. However, it appears that there is some confusion as to what constitutes a ‘repeated measure’ and what does not. This paper clarifies this confusion by defining characteristics a design may possess when the subjects are assessed multiple times. These characteristics are delineated based on whether the subjects are the experimental units, if the experimental units are measured repeatedly or not, and which randomisation strategy is used. Experimental designs can then be classified using these characteristics and visualised using Hasse diagrams.

%\includegraphics{figs/cas-grabs.pdf}
%\item Research highlights item 1
%\item Research highlights item 2
%\item Research highlights item 3

\begin{flushleft}
\end{flushleft}

\section{Introduction}

Many experiments involving human participants (such as in clinical trials) or animals (in pre-clinical research) require taking multiple measurements on each experimental subject. For example, in clinical trials the patients’ response to a treatment can be assessed by taking repeated measurements over time, and in animal behavioural experiments each animal may be assessed three times per task to account for the within-animal variability. In this paper we use the term ‘subject’ to refer to either humans or animals. A benefit of taking multiple measurements on each experimental subject is that we can potentially gain more information from each subject than would have been the case if they were only assessed once.

Although the term ‘repeated measures design’ is routinely used to describe designs where each subject is assessed multiple times, there appears to be some confusion in the literature as to what specifically constitutes a repeated measures design and what does not. \citet{bate2014} define seven ways that an animal can be measured more than once, of which the ‘repeated measures design’ is just one of these scenarios. A second scenario, where animals receive a sequence of treatments over time, they define as a ‘cross-over’ design, see \citet{jones2003}. \citet{ruxton2011} similarly use the ‘cross-over’ term, but also refer to such designs as ‘within-subjects’ or ‘repeated measures’. \citet{piepho2004} reserve the term ‘repeated measures’ for the case where units of material are measured repeatedly, perhaps over time or position, in a non-random order. \citet{brien2009} include cross-over designs within this class, reserving ‘longitudinal’ for the case where multiple measurements are taken on some unit of experimental material in a non-random order. They remove the constraint that the order of the repeated times/positions cannot be randomised, stating only that other factors cannot be randomised to them. 

Despite widespread use of the term ‘repeated measures design’, it is often applied inconsistently to fundamentally different types of experimental designs. Consider designs in which subjects are measured repeatedly over time, designs in which subjects receive multiple treatments in a non-random order, and cross-over designs, in which treatment sequences are randomised to subjects. All of these different designs are frequently described as ‘repeated measures designs’, despite having different experimental units, randomisation strategies, and modelling approaches. This lack of clarity can lead to inappropriate analyses. For example, if the subjects are measured repeatedly over time, then the researcher may include a Time factor as an additional factor in a single strata ‘analysis of variance’-based analysis, effectively assuming independence between the within-subject responses. This implies that the fixed effects in the model will not be assessed against the standard errors, leading to potentially misleading inferences, see \citet{littell2000}.  

In this paper we will describe a set of design characteristics, based on whether the experimental subjects are the experimental units, if the experimental units are measured repeatedly or not, and which randomisation strategy was employed. Different designs can then be defined based on the characteristics they possess, leading to a better understanding of the design and hence the correct analysis strategy. To further clarify the different designs and their properties, each design type will be visualized using a Hasse Diagram \citep{bate2016a, bate2016b}.

\section{Background definitions and concepts}

When differentiating experimental designs that involve measuring the subjects multiple times, it is important to identify the observational and experimental units. We follow the approach of \citet{bate2024clare} and define an experimental unit as the unit that independently receives a treatment such that the findings obtained from one experimental unit are unrelated to the findings obtained from any other. Note that under this definition a design may have more than one type of experimental unit, for example in a split-plot design. An observational unit is the smallest unit on which a response is measured \citep[p. 8]{bailey2008}. For example, in a clinical trial, where each subject provides a single blood sample which is separated into three aliquots for analysis. The subject is typically the experimental unit for the treatment being investigated, as each subject is randomised to a treatment group, whereas each aliquot is an observational unit on which a response is measured.

All experimental designs involve the use of experimental factors, and we follow \citet{bate2016a} and categorise them as ‘factors of interest’ to the researcher (i.e., treatment factors), factors ‘inherent’ to the structure of the experiment (i.e., a factor indexing the order of testing), factors that define the experimental and observational units (if not already identified as inherent) and factors associated with additional sources of variability (i.e., blocking factors). In this paper factor names are differentiated from the corresponding effects by using initial capital letters for the factor names.

To differentiate designs where the subjects are assessed multiple times, two further factor types are required. The Subject factor is an inherent factor whose levels index the units of experimental material, in this paper either humans or animals, that cannot be manipulated or reassigned by the experimenter, but which define the units on which measurements are repeatedly taken. The Repeated factor is either an inherent factor or a factor of interest and indexes the multiple observations taken on each level of the Subject factor. Repeated factors include Time (when subjects are measured repeatedly over time) but can be the factor that defines the observational units (for example, in a nested design). Of crucial importance, as we shall see below, is whether or not the levels of the Repeated factor are involved in the randomisation. The presence of a Repeated factor in the design does not automatically imply a lack of randomisation. Instead, it signals that additional modelling decisions may be required to account for any correlation induced if no randomisation was performed.

There are four ways that factors can be related to each other, either nested, fully crossed, partially crossed or equivalent. Factor $F_{\alpha}$ is said to be nested within factor $F_{\beta}$, denoted by $F_{\alpha}(F_{\beta})$, if within the experimental design each level of factor $F_{\alpha}$ occurs with one and only one level of factor $F_{\beta}$. $F_{\alpha}$ is said to be finer than $F_{\beta}$ and $F_{\beta}$ is said to be coarser than $F_{\alpha}$. Two factors $F_{\alpha}$ and $F_{\beta}$ are said to be fully crossed if all the levels of factor $F_{\alpha}$ occur with all the levels of factor $F_{\beta}$, and vice versa, see \citet[p. 180]{montgomery2017}. Two factors $F_{\alpha}$ and $F_{\beta}$ are said to be partially crossed if they are not fully crossed, but at least one level of factor $F_{\alpha}$ occurs at more than one level of factor $F_{\beta}$, and vice versa. For example, partial crossing within the design arises when subjects are measured at different subsets of time points or when natural constraints prevent all combinations of factors from occurring. This definition of crossing is simplified from the four types described by \citet{bailey2004}. Two factors are defined as equivalent within the design if for every occurrence of a level of factor $F_{\alpha}$ within the design, the same level of Factor $F_{\beta}$ occurs, and vice versa. In other words, the two factors are the same apart from the names of their levels (\citet{tjur1984} and \citet[p. 170]{bailey2008}. 

\citet{bate2016a} and \citet{bate2016b} describe an approach to identify the structure of the experimental design and then use this knowledge, alongside the randomisation performed, to construct a mixed model that can be used to analyse the experimental data. The approach involves identifying the experimental factors and pairwise relationships between them (either nested, crossed, partially crossed or equivalent). This information is summarised in the layout structure (LS). The LS consists of (i) a list of ‘structural’ objects, comprising of the factors that define the experimental design and any generalised factors (whose levels correspond to combinations of any crossed factors) that are implied by the structure of the experimental design and (ii) a description of the nested and crossed relationships between the factors and generalised factors, as defined by the experimental design. 

The LS, in combination with the randomisation performed, can be used to generate the restricted layout structure (RLS). This structure consists of ‘randomisation’ objects that are involved in the randomisation and correspond to objects in the LS. This list can then be used to determine the terms included in a mixed model that is 'best justified' by the randomisation \citep{goos2012}.

The layout (or restricted layout) structure can be visualised by a generalised Hasse diagram. These diagrams consist of the structural (or randomisation) objects corresponding to factors/generalised factors that are present in the LS and RLS, with solid lines connecting them to indicate nesting relationships and dotted lines connecting them to indicate partial crossing. Additionally, factors corresponding to random effects are underlined to differentiate them from factors corresponding to fixed effects. In the diagram of the RLS the individual randomisations can be illustrated by including an arrow whose direction indicates that the levels of one factor are randomly assigned to the levels of another, see \citet{bate2016b}. Also included on the diagram, as a  label next to each object, is the number of levels of the object and the corresponding 'structural' degrees of freedom calculated by the subtraction method \citep{brien1992}. The R package \texttt{hassediagrams} \citep{michaelides2026} has been developed to generate these Hasse diagrams (more in Appendix \ref{appendixA}).

\section{Design characteristics} \label{sec:characteristics}

Experimental designs that involve repeatedly measuring the subjects can be differentiated by considering certain characteristics they possess; these characteristics, and features intrinsic to them, are defined in Table~\ref{tab:design-characteristics}. For example, a design that involves subjects being randomly assigned to different sequences of treatments over multiple test-periods possess the ‘cross-over’ characteristic and can be defined as a ‘cross-over design’. Note it is possible for a design to possess more than one characteristic. For example, if the subjects in a cross-over design are repeatedly measured at multiple timepoints within each test-period, then the design also possesses the repeated measures characteristic and hence is defined as a ‘repeated-measures cross-over design’ (see Section \ref{subsec:drmd}).

\begin{ThreePartTable}
\tiny

\begin{TableNotes}
\tiny
\item[*] Characteristic does not depend upon definition of experimental unit.
\item[\textdagger] Core design is defined in Section \ref{subsec:rmdesign}.
\item[\textdaggerdbl] The $\text{\textasciicircum}$ symbol defines the generalised factor, i.e., $\text{A\textasciicircum B}$ is the generalised factor whose levels are combinations of A and B, see \citet{bailey1996} and \citet{tjur1984}.
\end{TableNotes}

\begin{longtable}{|L{1.6cm}|L{1.9cm}|L{1.7cm}|L{2.1cm}|L{2.1cm}|L{1.9cm}|L{2.0cm}|}
\caption{Summary of design characteristics when the levels of the Subject factor are measured repeatedly.}
\label{tab:design-characteristics}\\

\hline
\textbf{Characteristic} & \textbf{Example} & \textbf{Subject factor} & \textbf{Repeated factor} & \textbf{Relationship between Subject factor and Repeated factor} & \textbf{Experimental Units} & \textbf{Randomisation} \\
\hline
\endfirsthead

Repeated measures &
Human subjects receive a single treatment then are measured repeatedly over time &
Observational unit of the Core design\tnote{\textdagger} &
Factor not involved in randomisation that indexes repeated measurements &
Subject factor and Repeated factor crossed &
Indexed by a factor in the Core design &
Randomisation involving Core design factors; Repeated factor not involved in randomisation \\
\hline

Within-subjects &
Animals receive all treatments in the same non-random order &
Factor that receive multiple treatments &
Factor, confounded with Treatment, that indexes measurements across subjects &
Subject factor and Repeated factor crossed &
Indexed by Subject\textasciicircum Repeated generalised factor\tnote{\textdaggerdbl} &
Either (i) none or (ii) levels of factor(s) of interest randomised to Repeated factor levels \\
\hline

Block &
All treatments applied at random to positions in the subject &
Factor that defines the blocks &
Factor that defines within-subject `plots' nested within Subject factor &
Repeated factor nested within Subject factor &
Indexed by levels of the Repeated factor &
Factors of interest randomised to the Repeated factor levels, separately for each subject \\
\hline

Split-plot &
Multi-Reader ROC studies. Humans assess using multiple modalities but results are assessed by only one reader per human &
Factor that defines the `whole-plots' &
Factor that indexes the within-subject sub-plots of the split-plot design &
Repeated factor nested within Subject factor &
Subject for the `whole-plot' treatments and Repeated factor for the `sub-plot' treatments &
Whole-plot treatments are randomised to subjects, sub-plot treatments are randomised to the levels of the Repeated factor separately for each whole-plot \\
\hline

Cross-over &
Animals receive different sequences of treatments over time &
Inherent factor that defines the patients or animal &
Factor not involved in randomisation that indexes repeated measurements, i.e., Test period &
Subject factor and Repeated factor crossed &
Indexed by Subject\textasciicircum Repeated generalised factor\tnote{\textdaggerdbl} &
Planned sequences of levels of factor(s) of interest randomised to subjects \\
\hline

Nested &
Blood sample from each experimental unit assayed in triplicate &
Inherent factor that nests the observational unit factor &
Factor/generalised factor, nested within Subject factor, that indexes observations on each subject &
Repeated factor nested within Subject factor &
N/A\tnote{*} &
Levels of the Repeated factor randomised for each subject \\
\hline

\insertTableNotes

\end{longtable}
\end{ThreePartTable}

\section{Designs that involve repeatedly measuring subjects} \label{sec:designs_rms}

In this section classes of designs that involve repeatedly measuring the subjects are defined based on the characteristics described in Table~\ref{tab:design-characteristics}. Without loss of generality only designs with one characteristic will be considered. Examples of more complex designs, displaying multiple characteristics, are given in Section \ref{sec:morecomplexdesigns}. The number of ‘factors of interest’ is kept to a minimum; hence designs involving factorial treatment structures will not be considered, although the methods easily generalise to encompass such designs.

\subsection{Repeated measures design} \label{subsec:rmdesign}

A design is defined as a repeated measures design if it possesses the repeated measures characteristic, where multiple measurements are taken on the subjects in a non-random order. For example, responses are measured on each subject at specific time points or named brain regions. A repeated measures design consists of (i) a Core design and (ii) a Repeated factor, where the Repeated factor is crossed/partially crossed with all factors/generalised factors in the Core design. 

The Core design can be any experimental design, including a repeated measures design, and is the design that is assessed repeatedly. The structure of the Core design (the factor levels and their inter-relationships) does not change as repeated measurements are taken. The observational unit of the Core design (indexed by the factor OU$_{\text{core}}$) is the unit that is measured repeatedly. In this paper, OU$_{\text{core}}$ is usually the Subject factor.

The Repeated factor, which is often a ‘factor of interest’, indexes (across all subjects) the repeated measurements. The levels of the Repeated factor are not involved in the randomisation and it is crossed or partially crossed with OU$_{\text{core}}$ (and hence is crossed/partially crossed with all other factors in the Core design). Effectively, the levels of the Repeated factor index (across all subjects) the repeated measurements taken.

The factor corresponding to the experimental unit will always reside within the Core design. Repeated factor(s) are not involved in the randomisation, so any randomisation of the levels of the factors in the design will occur within the Core design. 

A schematic illustration of the Hasse diagram of the LS of the repeated measures design is given in Figure~\ref{fig:core}.

\begin{center}
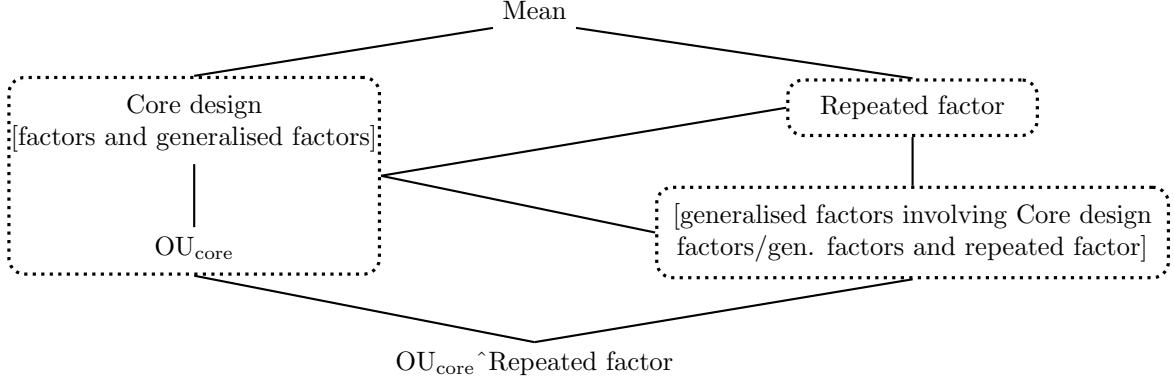

\begin{minipage}{\textwidth}
\centering

\begin{tikzpicture}[
  box/.style={draw, dotted, very thick, rounded corners=6pt, align=center, inner sep=6pt},
  line/.style={thick}
]

% Center axis
\node (mean) at (0,4.2) {Mean};

% Spread symmetrically
\node[box, minimum width=4.9cm, minimum height=2.6cm] (core) at (-4.5,2) {};

\node (coretitle) at (-4.5,2.7) {
\begin{tabular}{c}
Core design\\
{[factors and generalised factors]}
\end{tabular}
};

\node (oucore) at (-4.5,1.05) {OU$_{\text{core}}$};

\draw[line] (coretitle.south) -- (oucore.north);

\node[box, minimum width=3.3cm, minimum height=0.7cm] (rep) at (5,2.9)
{Repeated factor};

\node[box, minimum width=6.0cm, minimum height=1.0cm] (gen) at (5,1.25)
{{[generalised factors involving Core design}\\
{factors/gen. factors and repeated factor]}};

% Shared bottom point (exact center)
\coordinate (bottompoint) at (0,-0.2);
\node[below=0pt of bottompoint] (bottom)
{OU$_{\text{core}}$\textasciicircum Repeated factor};

% Top connections (balanced angles)
\draw[line] (mean.south west) -- (core.north);
\draw[line] (mean.south east) -- (rep.north);

% Middle structure
\draw[line] (core.east) -- (rep.west);
\draw[line] (core.east) -- (gen.west);
\draw[line] (rep.south) -- (gen.north);

% Bottom convergence (perfectly centered)
\draw[line] (core.south) -- (bottompoint);
\draw[line] (gen.south) -- (bottompoint);

\end{tikzpicture}

\captionof{figure}{Schematic Hasse diagram of the LS of a repeated measures design.}
\label{fig:core}

\end{minipage}
\end{center}

An approach to define the mixed model based on the structure of the experimental design and the randomisation performed was described in \citet{bate2016b}. The mixed model can be represented in matrix notation as:
\[
y = X\beta + Z\zeta + \varepsilon,
\]
where $y$ is a vector of observations, $X$ and $Z$ are the design matrices for the fixed and random categorical factors respectively, $\beta$ is the vector of fixed effect parameters, $\zeta$ is the vector of random parameters, with a variance components structure, and $\varepsilon$ is a vector of IID random effects.

Such models can be applied when all terms have been ‘involved in the randomisation’ and hence the mixed model is obtained as a direct result of the randomisation. Known as the linear predictor model, it is appropriate for the analysis as it is the best model justified by the randomisation \citep{goos2012}.

However, in the experiments discussed in this paper the Repeated factor is not involved in the randomisation and hence neither the Repeated factor nor any interactions involving the Repeated factor can be included in a mixed model following the strategy of \citet{bate2016b}. As the randomisation does not support the inclusion of such terms in the model, then to include the Repeated factor (and any interactions) in the final mixed model, the correlation between observations, implied by the lack of randomisation, should be modelled. For example, a repeated measures mixed model (MMRM) approach could be used so that the Repeated factor, and any interactions involving the Repeated factor, can be included in the modelling process \citep{wang2023, Rmmrm}.

The approach described in \citet{bate2016a} and \citet{bate2016b} therefore requires some additional steps.

\begin{description}

\item[\textbf{Stage 1:}]
To begin with the RLS is constructed as described in \citet{bate2016b}.

\item[\textbf{Stage 2:}]
The observational unit of the Core design (OU$_{\text{core}}$) is included in the RLS, if not already present.

\item[\textbf{Stage 3:}]
The RLS is then extended by including the Repeated factor and any generalised factors between the factors/generalised factors of the core design and the Repeated factor.

\end{description}

\subsection*{Example 1. ACE inhibitor clinical trial} \label{subsec:example1}

\noindent \citet{senn2000} describe a multi-centre randomised double-blind placebo controlled clinical trial to assess the effect of an intervention of early stage ACE inhibitor on renal disease in insulin dependent diabetic patients. The patients (factor: Subject) were randomly assigned to either drug or placebo arms (factor: Drug) and were assessed at 5 equally spaced time points over 24 months (factor: Time). Assuming each centre (factor: Centre) included patients allocated to both drugs, the LS for the Core design, as defined using the approach described in \citet{bate2016a}, included Drug, Centre, $\text{Drug\textasciicircum Centre}$ and Subject. If treatments were randomly assigned to patients separately for each centre, then using the arrow notation of \citet{bate2016b} the randomisation can be defined using two randomisation objects and a single randomisation arrow:
\[
\text{Drug} \;\rightarrow\; \text{Subject}[\text{Centre}].
\]

\begin{figure}
\centering
\includegraphics[width=0.9\textwidth]{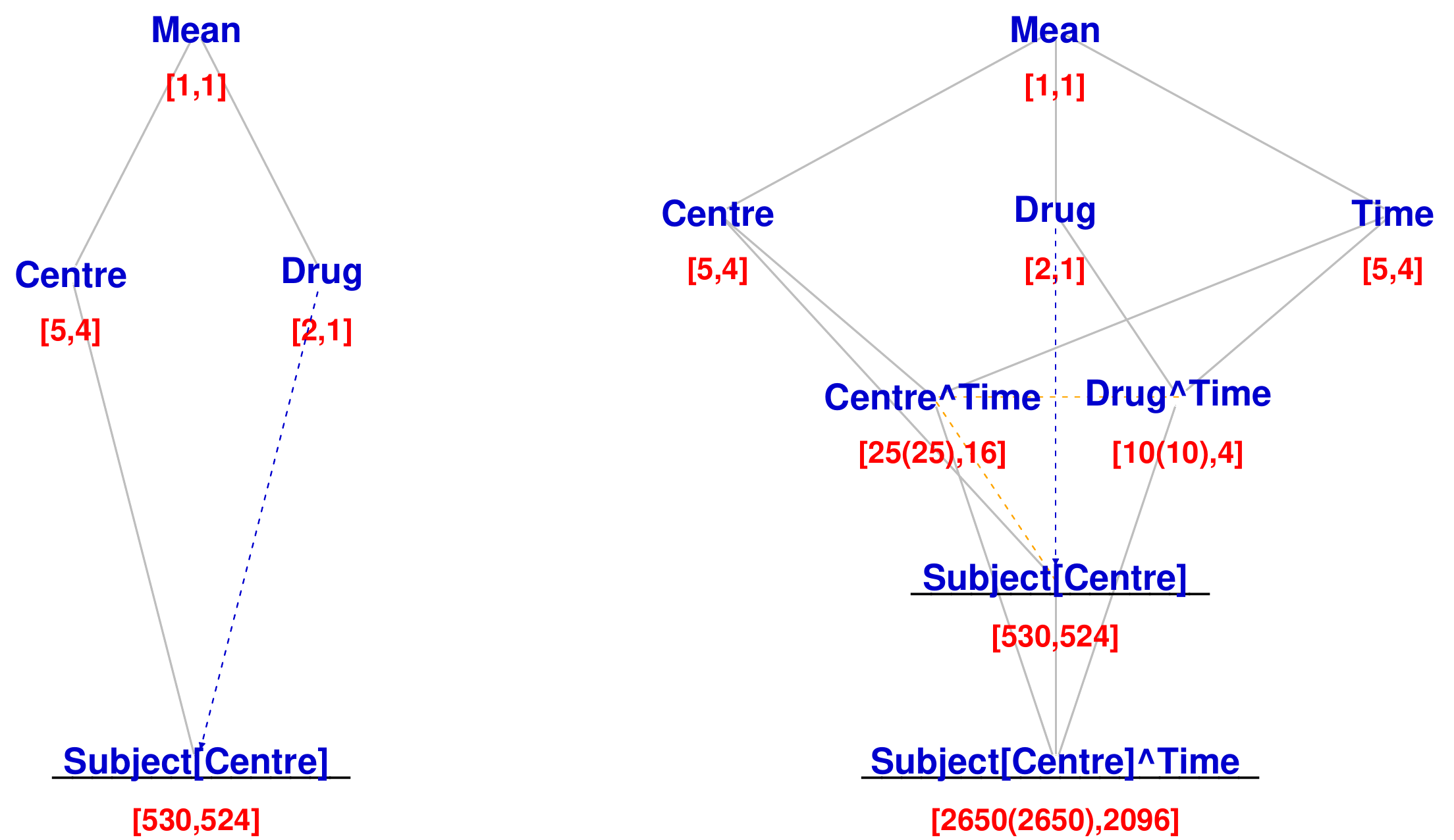}
\caption{Hasse diagram of (a) RLS of the Core design and (b) RLS of the repeated measures design for the ACE inhibitor clinical trial.}
\label{fig:core_repeated}
\end{figure}

The RLS of the Core design in \citet{bate2016b} includes the randomisation objects Drug, Centre and Subject[Centre] and can be visualised using a generalised Hasse diagram (panel (a) of Figure~\ref{fig:core_repeated}).

The levels of the Time factor index the repeated measurements on each patient and are not randomised, so Time is a Repeated factor and hence following the approach discussed above, the RLS of the repeated measures design includes the randomisation objects derived from the Core design, the Repeated factor Time and any generalised factors involving Time and randomisation objects from the Core design. The generalised Hasse diagram of the repeated measures design is given in panel (b) of Figure~\ref{fig:core_repeated}. Note the figures in brackets are the total number of possible combinations of the factors and the number preceding them is the number present in the design \citep{bate2016a}. The dotted line implies that the generalised factors $\text{Centre\textasciicircum Time}$ and $\text{Drug\textasciicircum Time}$ are partially crossed with each other.

\subsection{Within-subject design} \label{subsec:wsd}

A design that possesses the within-subject characteristic is defined as a within-subject design. All the subjects receive the same series of interventions, one per ‘test-period’, in a non-random order. These designs share similarities with the repeated measures designs as the factor of interest (that defines the levels of the intervention) is equivalent to the factor that defines the test-periods (factor: Period) and hence is itself a Repeated factor.

The Subject factor will be crossed (or partially crossed) with the Repeated factor as each of the subjects will be subjected to multiple different interventions. In within-subject designs, the experimental units are not the subjects themselves but the combinations of subjects and levels of the Repeated factor, reflecting the fact that each subject receives multiple interventions.

A schematic illustration of the Hasse diagram of the LS of a design with the within-subject characteristic is given in Figure~\ref{fig:corewithin}.

\begin{center}
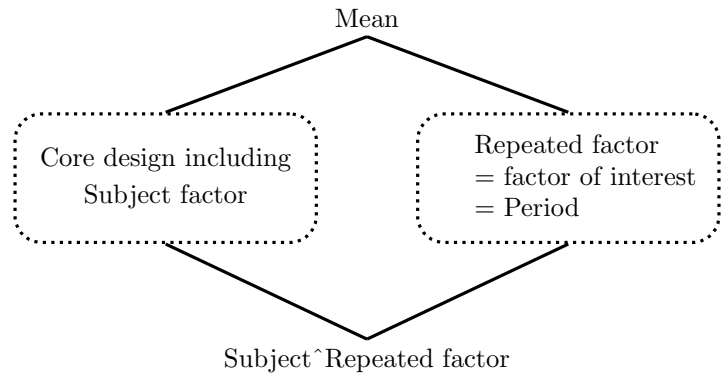

\begin{minipage}{\textwidth}
\centering

\begin{tikzpicture}[
  node distance=2.2cm and 1.2cm,
  box/.style={
    draw,
    dotted,
    very thick,
    rounded corners=10pt,
    align=center,
    text width=3.75cm,
    minimum height=1.7cm
  }
]

\node (mean) {\normalsize Mean};

\node[box, below left=1.0cm and 0.1cm of mean] (core)
  {\normalsize Core design including\\[2pt]\normalsize Subject factor};

\node[box, below right=1.0cm and 0.1cm of mean] (rep)
{
\normalsize
\begin{tabular}{l l}
 & Repeated factor \\
& = factor of interest \\
& = Period
\end{tabular}
};

\node[below=4.0cm of mean] (subperiod)
  {\normalsize Subject\textasciicircum Repeated factor};

\draw[very thick] (mean.south) -- (core.north);
\draw[very thick] (mean.south) -- (rep.north);

\draw[very thick] (core.south) -- (subperiod.north);
\draw[very thick] (rep.south) -- (subperiod.north);

\end{tikzpicture}

\captionof{figure}{Schematic Hasse diagram of the LS of a design with the within-subject characteristic.}
\label{fig:corewithin}

\end{minipage}
\end{center}

As the levels of the factor of interest (the Repeated factor) are administered to the subjects in a non-random order, so the within-subject results may be correlated. This correlation structure needs to be considered and implies a repeated measures mixed model approach should be used to assess the data generated.

\subsection*{Example 2. Dog telemetry study} \label{subsec:example2}

\begin{figure}
\centering
\includegraphics[width=0.9\textwidth]{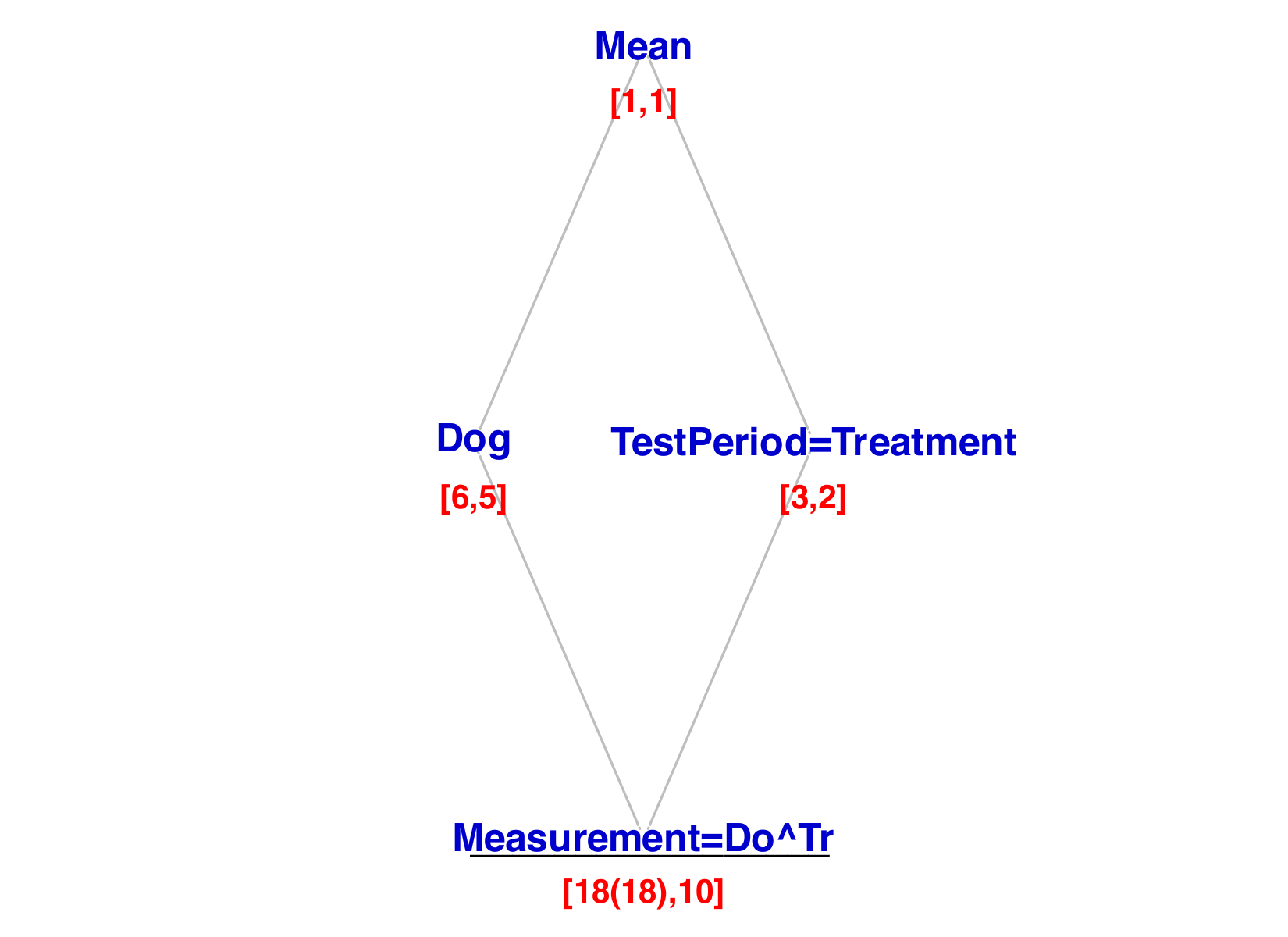}
\caption{Hasse diagram of the LS for a six dog, four treatment six dog telemetry study using a within-subject dose-escalation design.}
\label{fig:within}
\end{figure}

\noindent When assessing the safety profile of novel pharmaceutical compounds and their effect on the heart, a dog telemetry study is commonly used. Dogs (factor: Dog), usually 4 or 6 animals per study, receive three doses of the compound and control (factor: Treatment) over time, one per test-period (factor: Period). To reduce the risk of side effects, a dose-escalation design is often employed, where each animal first receives control, followed by the low, then the middle and then high dose. Progression to the higher dose only occurs if no adverse effects are observed at the current dose. 
In this type of study, the Dog factor is crossed with the Treatment factor and, as the order of treatment administration is non-random, so the within-animal correlation needs to be accounted for in the statistical analysis. The Hasse diagram of the LS of the six dog design is given in Figure~\ref{fig:within}, where it is assumed that a single measurement is taken on each dog per treatment period (factor: Measurement). 

\subsection{Block design}

Consider an experiment where each level of the Subject factor receives multiple levels of the factor of interest (i.e., treatments), but the allocation of treatments within each subject is randomised separately for each subject. Because of the randomisation of the design (i.e., a separate randomisation for each subject) the levels of the Repeated factor, that define the multiple measurements per subject, are nested within the Subject factor. Such a design possesses the block characteristic, and hence, is defined as a block design, where the subjects define the blocks and the levels of the Repeated factor correspond to the plots within blocks (factor: Plot). As multiple treatments are administered to plots within each subject, the treatment factor(s) will be crossed/partially crossed with Subject factor. Using the arrow notation of \citet{bate2016b} this randomisation can be written using two randomisation objects as:
\begin{equation*}
\text{Treatment} \to \text{Plot[Subject]}.
\end{equation*}

A schematic illustration of the Hasse diagram of the LS of a block design, where the subjects are a blocking factor, is given in Figure~\ref{fig:blocktikz}. Note if each treatment is administered once to each subject, then the Plot factor will be equivalent to the $\text{Subject\textasciicircum Treatment(s)}$ generalised factor.

\begin{center}
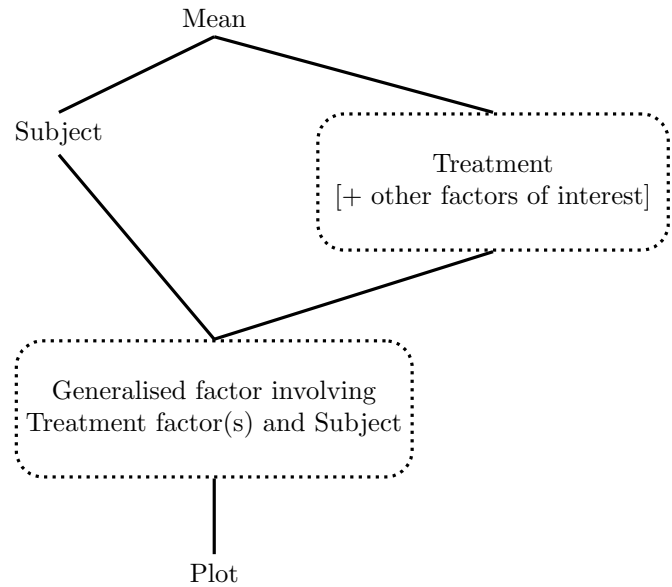

\begin{minipage}{\textwidth}
\centering

\begin{tikzpicture}[
  node distance=2.2cm and 1.5cm,
  box/.style={
    draw,
    dotted,
    very thick,
    rounded corners=10pt,
    align=center,
    text width=4.4cm,
    minimum height=1.8cm
  }
]

% Top node
\node (mean) {\normalsize {Mean}};

% Left node (Subject)
\node[below left=1.0cm and 0.8cm of mean] (subject)
  {\normalsize {Subject}};

% Right node (Treatment box)
\node[box, below right=1.0cm and 0.8cm of mean] (treat)
{
\normalsize
{Treatment}\\
{[+ other factors of interest]}
};

% Middle node (generalised factor)
\node[box, below=4.0cm of mean, text width=5cm] (gen)
{
\normalsize
{Generalised factor involving}\\
{Treatment factor(s) and Subject}
};

% Bottom node (Plot)
\node[below=1cm of gen] (plot)
  {\normalsize Plot};

% Connections
\draw[very thick] (mean.south) -- (subject.north);
\draw[very thick] (mean.south) -- (treat.north);

\draw[very thick] (subject.south) -- (gen.north);
\draw[very thick] (treat.south) -- (gen.north);

\draw[very thick] (gen.south) -- (plot.north);

\end{tikzpicture}

\captionof{figure}{Schematic Hasse diagram for the LS of a block design, where the Subject factor indexes the blocks and the Repeated factor defines the plots.}
\label{fig:blocktikz}

\end{minipage}
\end{center}

As the treatments are randomised independently to plots within each subject, treatment comparisons can be assumed to be independent, and hence, the statistical analysis can be performed using an ANOVA or mixed model approach, while accounting for between-subject variability through inclusion of a Subject blocking factor in the statistical model. The treatments can therefore to be assessed ‘within-subject’. However, the randomisation does not allow the assessment of any interactions involving the treatment and Subject factors.

\subsection*{Example 3. Insect repellent trial}

\begin{figure}
\centering
\includegraphics[width=0.9\textwidth]{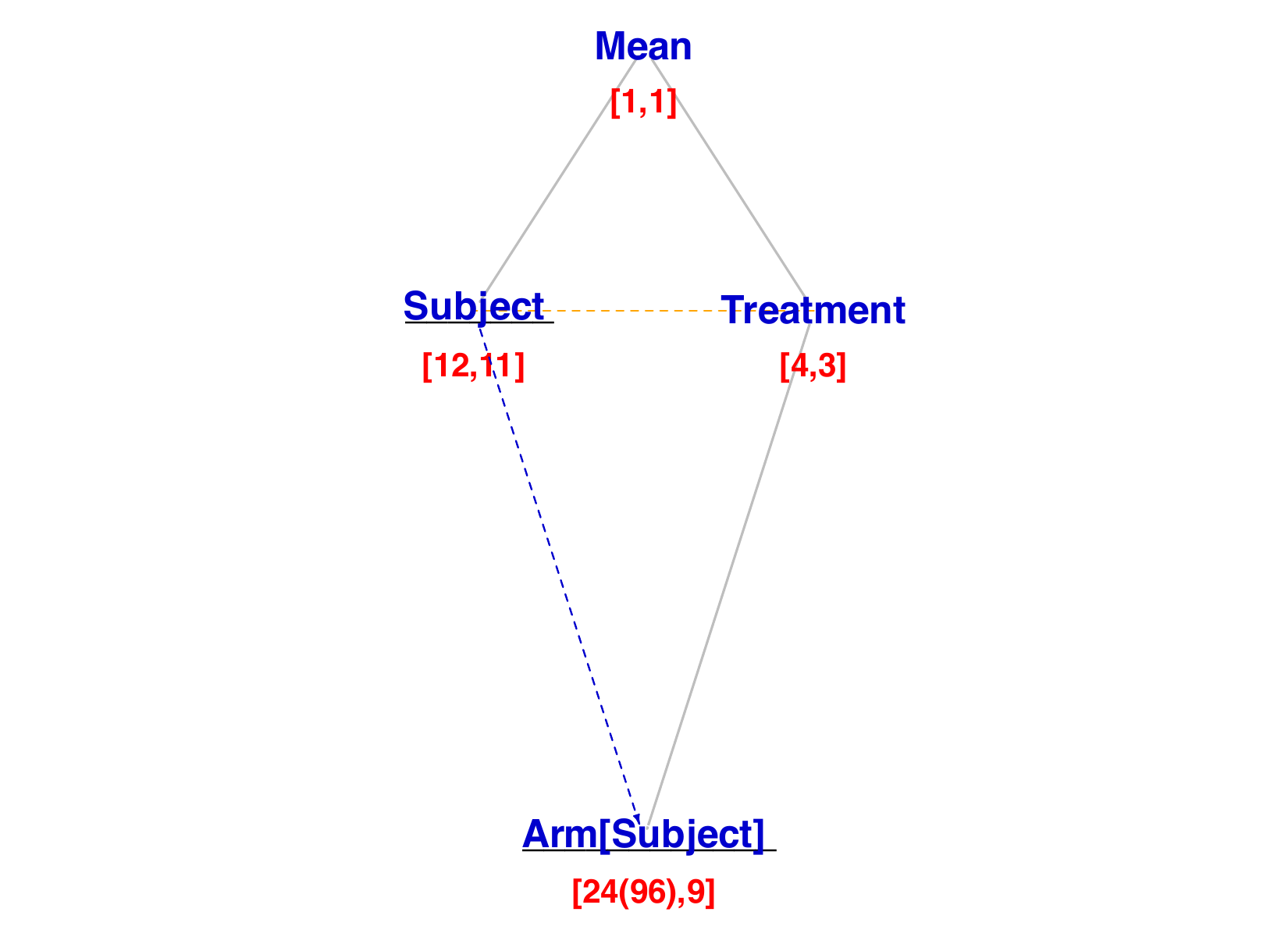}
\caption{Hasse diagram of the RLS, including randomisation arrow, for the insect repellent trial.}
\label{fig:block}
\end{figure}

\noindent \citet{bailey2008} describes an experiment to assess the effect of insect repellent on twelve volunteers. As people are known to attract insects to different degrees, each volunteer (factor: Subject) receives a different insect repellent (factor: Treatment) on each arm (Repeated factor: Arm), thus allowing a within-person assessment of the effect of the repellents. Effectively the design is a block design, blocking by person. Assuming 4 repellents were tested then it is sensible to use a balanced incomplete block design for 4 treatments with two plots per block, i.e., arms per person, (replicated twice). The design, in non-randomised order, is given in Table~\ref{tab:bibd}. Assuming the randomisation performed was:
\begin{equation*}
\text{Treatment} \to \text{Arm[Subject]},
\end{equation*}
the Hasse Diagram of the RLS is given in Figure~\ref{fig:block}.

\begin{center}
\captionof{table}{Balanced incomplete block design for 4 treatments, 2 treatments per block.}
\label{tab:bibd}
\begin{tabular}{|c|c|*{12}{c|}}
\hline
\multicolumn{2}{|c|}{} & \multicolumn{12}{c|}{Subject} \\
\hline
\multicolumn{2}{|c|}{} 
& 1 & 2 & 3 & 4 & 5 & 6 & 7 & 8 & 9 & 10 & 11 & 12 \\
\hline
\multirow{2}{*}{\begin{tabular}{c}Subject\\Arm\end{tabular}}
& Left  
& A & A & A & B & B & C & A & A & A & B & B & C \\
\cline{2-14}
& Right 
& B & C & D & C & D & D & B & C & D & C & D & D \\
\hline
\end{tabular}
\end{center}

\subsection{Split-plot design}

Consider a split-plot design where the whole-plot treatments are randomly assigned to subjects (or whole-plots) and, separately for each subject, the sub-plot treatments are allocated randomly within each subject. The two types of experimental unit in the split-plot design correspond to (i) the subjects (whole-plot treatments are assigned to subjects) and (ii) the sub-plots (sub-plot treatments are randomly assigned to sub-plots within subject). There are therefore two randomisations performed
\begin{equation*}
\text{Whole-plot treatment(s)} \to \text{Subject},
\end{equation*}
and 
\begin{equation*}
\text{Sub-plot treatment(s)} \to \text{Sub-plot(Subject)}.
\end{equation*}

In general, the whole-plot treatment factor(s) nest the Subject factor whereas the sub-plot treatment factor(s) are crossed with the Subject factor. As sub-plot treatments are randomly assigned to sub-plots within subjects, the sub-plot factor is a Repeated factor and it is nested within the Subject factor.

Due to the randomisation and the random factors (Whole-plot and Sub-plot) the data is usually analysed using either a mixed-effects ANOVA or mixed model approach.

\subsection*{Example 4. Coated implant assessment in rabbits} \label{subsec:example4}

\noindent An experiment was conducted to assess the effect of four types of coated implants (factor: Type) on bone formation in rabbits \citep[p. 119]{bate2014}. Four cylinders, one per implant type, were inserted into random positions (factor: Position) within the femoral condyles of 18 rabbits (factor: Rabbit). The rabbits were humanely killed at three time points post-surgery (factor: Time), six per time point. The design used was a split-plot design with Rabbit defining the whole plots, Time as the whole-plot treatment factor, Type of implant as the sub-plot treatment factor and Position defining the sub-plots. To begin the randomisation, rabbits were randomly assigned to the three time points
\begin{equation*}
\text{Time} \to \text{Rabbit},
\end{equation*}
then separately for each rabbit, the four types of implant were randomly assigned to positions within each rabbit
\begin{equation*}
\text{Type} \to \text{Position[Rabbit]}.
\end{equation*}

\begin{figure}
\centering
\includegraphics[width=0.9\textwidth]{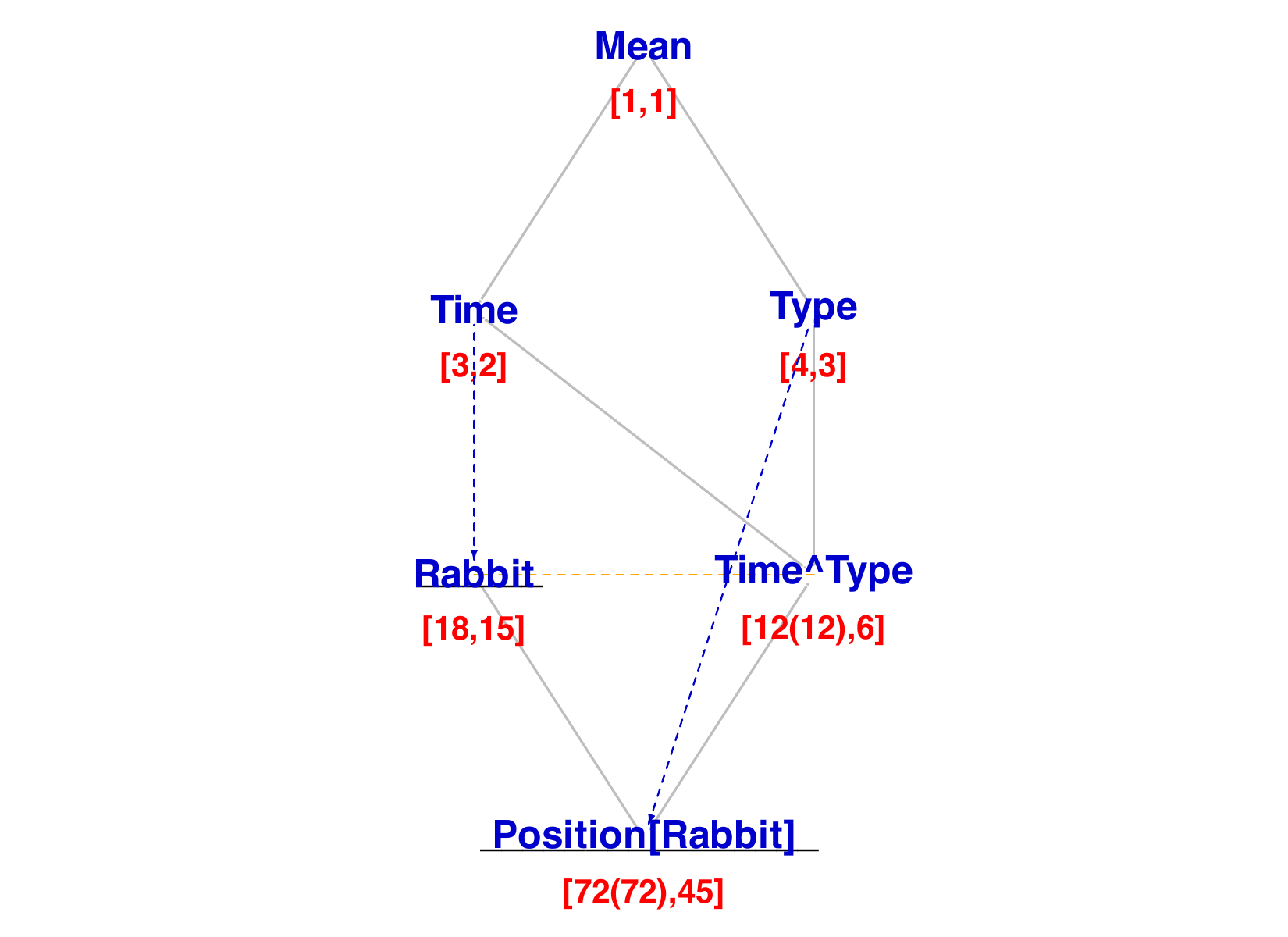}
\caption{Hasse diagram of the RLS for the coated implant assessment in rabbits.}
\label{fig:splitplot}
\end{figure}

Using the method described in \citet{bate2016b} Time, Type, Position, Rabbit and the $\text{Time\textasciicircum Type}$ generalised factor are included in the RLS. The Hasse diagram for the RLS, including the randomisation arrows is given in Figure~\ref{fig:splitplot}.

\subsection{Cross-over design} \label{subsec:crossoverdesign}

A cross-over design is a design that possess the cross-over characteristic: i.e., the subjects (factor: Subject) receive a sequence of different levels of a ‘treatment’ factor (factor: Treatment) over time, one level per test-period (Repeated factor: Period). The treatment sequences (factor: Sequence) are different for each subject, or alternatively subjects are grouped together and each group receives the same sequence of treatments. 

Treatment sequences are selected in advance so that, when taken together, they have certain properties; for example each treatment precedes each other treatment equally often in the design. This implies the randomisation of treatments to subjects and test-periods is constrained so that the structure of the experimental design is preserved. Subjects are randomised to the treatment sequences and treatments are randomised to the treatment labels of the design.

The experimental units in such designs are the levels of the $\text{Subject\textasciicircum Period}$ generalised factor. This generalised factor often also indexes the observational units, unless subjects are measured repeatedly within a test-period, see Section \ref{subsec:drmd}. 

There are similarities between designs possessing the within-subject characteristic, described in Section \ref{subsec:wsd}, and those with the cross-over characteristic, as both involve each subject receiving a sequence of multiple treatments over the levels of the Repeated factor. However, the cross-over characteristic implies the treatment sequences that a subject receives are, in some sense, randomised; whereas the order is non-random and the same for all subjects in the within-subject design. 

The factors Treatment, Subject and Period are crossed (or partially crossed). Subject is either nested within Sequence or equivalent (if each subject receives a different sequence of treatments). A schematic illustration of the Hasse diagram of the LS of a cross-over design for the scenario where groups of subjects receive the same treatment sequence and are measured once per test period is given in Figure~\ref{fig:crossover}.

\begin{center}
\begin{minipage}{\textwidth}
\centering

\begin{tikzpicture}[
  box/.style={
    draw,
    dotted,
    very thick,
    rounded corners=8pt,
    align=center,
    minimum height=1.0cm
  }
]

% Nodes
\node (mean) at (0,0) {\normalsize Mean};

\node (seq) at (-4.4,-1.1) {\normalsize Sequence};
\node (period) at (0,-1.1) {\normalsize Period};
\node[box, text width=4.7cm] (treat) at (4.2,-1.35)
  {\normalsize Treatment\\[3pt]\normalsize [+ other factors of interest]};

\node (subseq) at (-4.4,-2.05) {\normalsize Subject(Sequence)};

\node[box, text width=4.8cm] (genperiod) at (2.35,-3.65)
  {\normalsize [Generalised factors involving\\[3pt]
   \normalsize Treatment factor(s) and Period]};

%\node[box, text width=5.8cm] (bottomgen) at (0,-5.5)
%  {\normalsize Subject(Sequence)\textasciicircum Period\textasciicircum Treatment};

% \node (obs) at (0,-7) {\normalsize Subject(Sequence)\textasciicircum Period\textasciicircum Treatment = Observational unit};
\node (obs) at (0,-6) {\normalsize Subject(Sequence)\textasciicircum Period\textasciicircum Treatment = Observational unit};

% Lines
\draw[very thick] (mean.south) -- (seq.north);
\draw[very thick] (mean.south) -- (period.north);
\draw[very thick] (mean.south) -- (treat.north);

\draw[very thick] (seq.south) -- (subseq.north);

\draw[very thick] (period.south) -- (genperiod.north);
\draw[very thick] (treat.south) -- (genperiod.north);

\draw[very thick] (subseq.south) -- (obs.north);
\draw[very thick] (genperiod.south) -- (obs.north);

%\draw[very thick] (bottomgen.south) -- (obs.north);

\end{tikzpicture}


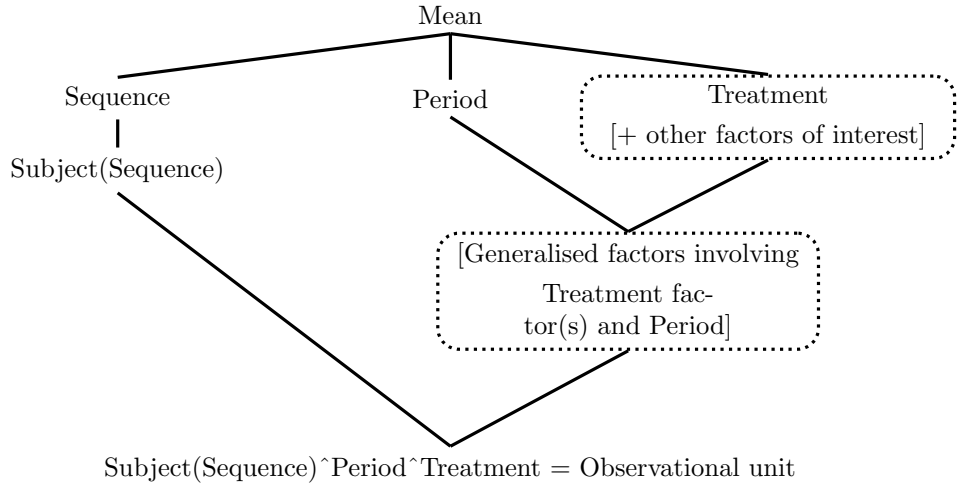
\captionof{figure}{Schematic Hasse diagram for the LS of a cross-over design.}
\label{fig:crossover}

\end{minipage}
\end{center}

As the treatment labels are randomised to the subject/period combinations, data is usually analysed using an ANOVA-based approach with fixed factors Treatment, Period and Subject. While traditional analyses of cross-over designs are often framed in terms of fixed-effect ANOVA models, the modelling of between- and within-subject variability can be performed using a mixed model approach with Subject as a random effect. Subject may also be modelled as a random effect when Sequence is included in the model as Subject is nested within Sequence. 

It should be noted there are similarities between designs possessing the block (by subject) and cross-over characteristics, especially when the researcher wishes to block in two directions and perhaps use a row-column block design. However, while the general structure of such designs may be the same, i.e., both could be based on Latin squares, the randomisation is different. In a cross-over design the order that the treatments are administered to the subjects is planned in advance, and hence the randomisation reflects this:
\begin{equation*}
\text{Sequence} \to \text{Subject}.
\end{equation*}
In a row-column block design the order of allocation of treatments to plots is randomised separately for rows (i.e., Period) and columns (i.e., Subject):
\begin{equation*}
\text{Treatment} \to \text{Subject $\otimes$ Period}.
\end{equation*}

The experimental units in a block (by subject) design are the individual plots, where plots are nested within subjects. The Plot factor is analogous to the levels of the generalised factor $\text{Subject\textasciicircum Period}$ in a cross-over design. 

\subsection*{Example 5. Intermittent claudication clinical trial}

\begin{figure}
\centering
\includegraphics[width=0.9\textwidth]{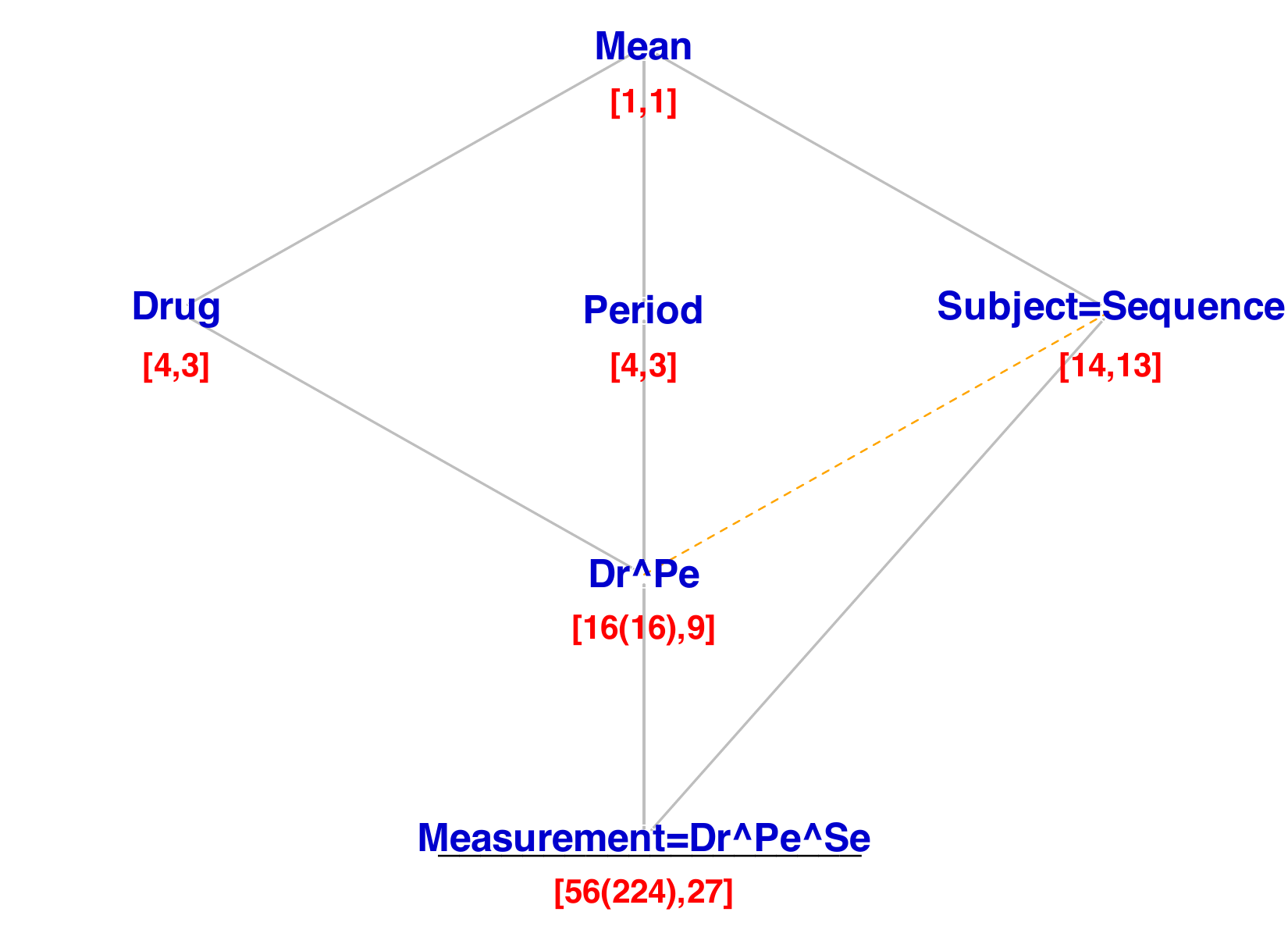}
\caption{Hasse diagram of the LS for intermittent claudication clinical trial.}
\label{fig:crossoverdesign}
\end{figure}

\noindent An experiment was conducted to assess the effect of four treatments on cardiac output in patients with intermittent claudication. The study involved 14 human subjects (factor: Subject), each of which received the four treatments (factor: Treatment) over an eight-week timeframe in a different sequence (factor: Sequence) for each patient. Each treatment was administered for one week followed by a one-week wash out period (factor: Period) and each patient was assessed once in each test period (factor: Measurement). The treatment regime is described in \citet[Table 1.1]{jones2003}. The Hasse diagram of the LS is given in Figure~\ref{fig:crossoverdesign}.

\subsection{Nested design}

Consider an experiment where the levels of the Subject factor are measured repeatedly, but there is no relationship between these measurements across subjects. Such a design is said to possess the nested characteristic and is defined as a nested design. For example, if a blood sample is taken from each patient (factor: Patient) in a clinical trial, split into three aliquots and each aliquot is assayed separately then there is no relationship between the first aliquots from patient 1 and the first aliquots from patient 2. The factor whose levels index the measurements (Repeated factor: Sub-sample) is nested within the Subject factor. 

In the simplest case, if individual subjects receive a single treatment and are then assessed multiple times (using a nested design) then the individual subjects that are the experimental units. The nested factor(s) are usually defined as random as there are no systematic differences between the levels of the nested factor(s). 

Many randomisations are possible with nested designs. \citet{white1975} describes three examples involving different ways that a three-way hierarchical nested design can be randomised. Usually though, the subjects will be randomised to the treatments. The order that the individual observations are processed may be randomised across or within subjects.

As nested designs involved a nested random factor(s), with perhaps a fixed factor that nests the random factor(s), the data can be analysed using a mixed-effects ANOVA or a mixed model approach. Given that the levels of the nested random factor are usually randomised in some way, then the correlations between the observations can be modelled using a variance components correlation structure.  

Note, if all the factors in the nested design are categorical, then the "lowest" nested factor defines the residual error, however if some of the factors are continuous then this term represents the pure error rather than the total error. These terms should potentially be kept separate on the Hasse diagram as it is recommended to use the pure error rather than the total error in any statistical tests \citep{gilmour2025}.

\subsection*{Example 6. Therapy assessment}

\begin{figure}
\centering
\includegraphics[width=0.9\textwidth]{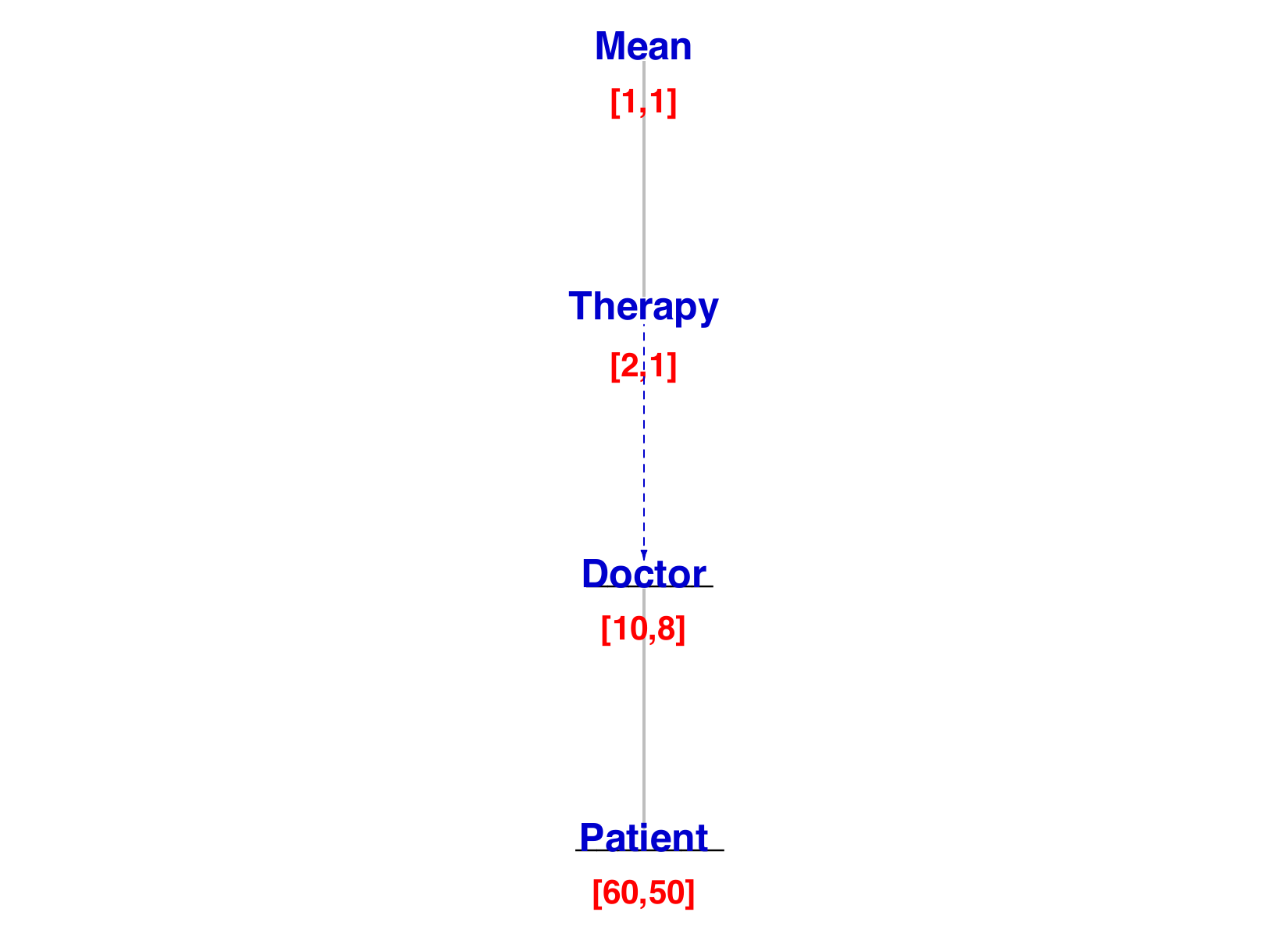}
\caption{Hasse diagram of the RLS, including randomisation arrow, for the therapy assessment trial.}
\label{fig:nested}
\end{figure}

\noindent An experiment was performed to assess the effect of two novel therapies (factor: Therapy) in human patients, see \citet{white1975}. Ten doctors (factor: Doctor) were randomised to one of two therapies, five doctors per therapy. These doctors then administered the therapy to six of their patients (Repeated factor: Patient).  The design used was a nested design, with the inherent factor Doctor nested within Therapy and Patient nested within Doctor (and Therapy). The Hasse diagram of the RLS, including the randomisation 
\begin{equation*}
\text{Therapy} \to \text{Doctor},
\end{equation*}
is given in Figure~\ref{fig:nested}. 

The doctors are the experimental units. While they are ‘measured’ six times (one per patient), the patients across doctors’ surgeries are not related, and hence this is an example of a nested design. 

\section{More complex designs involving repeated measures} \label{sec:morecomplexdesigns}

The characteristics described in Section \ref{sec:characteristics} highlight the different properties a design can possess when the subjects are measured repeatedly. Although the designs described in Section \ref{sec:designs_rms} only exhibit one of these characteristics, in practice a design can possess any number. In this section we consider some more complicated scenarios.

\subsection{Nested repeated measures design}

Consider a repeated measures design, as discussed in Section \ref{subsec:rmdesign}, where the observational units of the Core design ($\mathrm{OU}_{\mathrm{core}}$) are measured repeatedly and the Repeated factor indexes the levels of the repeated measures. Now assume that the combinations of the levels of the Repeated factor and $\mathrm{OU}_{\mathrm{core}}$ are themselves measured repeatedly, but there is no structure to these repeatedly measured responses. For example, the levels of the $\mathrm{OU}_{\mathrm{core}}\text{\textasciicircum}\mathrm{Repeated}$ factor are assayed in triplicate. In this case, the observational units are nested within the levels of the $\mathrm{OU}_{\mathrm{core}}\text{\textasciicircum}\mathrm{Repeated}$ factor and hence the design also possesses the nested characteristic of the design. Such a design is defined as a nested repeated measures design. A schematic Hasse diagram of the LS of a nested repeated measures design is given in Figure~\ref{fig:nestedrepeatedmeasures}.

\begin{center}
\begin{minipage}{\textwidth}
\centering

\begin{tikzpicture}[
  box/.style={draw, dotted, very thick, rounded corners=6pt, align=center, inner sep=6pt},
  line/.style={thick}
]

% Center axis
\node (mean) at (0,4.2) {Mean};

% Spread symmetrically
\node[box, minimum width=4.9cm, minimum height=2.6cm] (core) at (-4.5,2) {};

\node (coretitle) at (-4.5,2.7) {
\begin{tabular}{c}
Core design\\
{[factors and generalised factors]}
\end{tabular}
};

\node (oucore) at (-4.5,1.05) {OU$_{\text{core}}$};

\draw[line] (coretitle.south) -- (oucore.north);

\node[box, minimum width=3.3cm, minimum height=0.7cm] (rep) at (5,2.9)
{Repeated factor};

\node[box, minimum width=6.0cm, minimum height=1.0cm] (gen) at (5,1.25)
{{[generalised factors involving Core design}\\
{factors/gen. factors and repeated factor]}};

% Shared bottom point (exact center)
\coordinate (bottompoint) at (0,-0.2);
\node[below=0pt of bottompoint] (bottom)
{OU$_{\text{core}}$ \textasciicircum Repeated factor};

\node[box, minimum width=3.0cm, minimum height=0.7cm] (vbottom) at (0,-2)
{{[additional nested factors]}};

% Top connections (balanced angles)
\draw[line] (mean.south west) -- (core.north);
\draw[line] (mean.south east) -- (rep.north);

% Middle structure
\draw[line] (core.east) -- (rep.west);
\draw[line] (core.east) -- (gen.west);
\draw[line] (rep.south) -- (gen.north);

% Bottom convergence (perfectly centered)
\draw[line] (core.south) -- (bottompoint);
\draw[line] (gen.south) -- (bottompoint);

% final line
\draw[line] (bottom.south) -- (vbottom.north);

\end{tikzpicture}


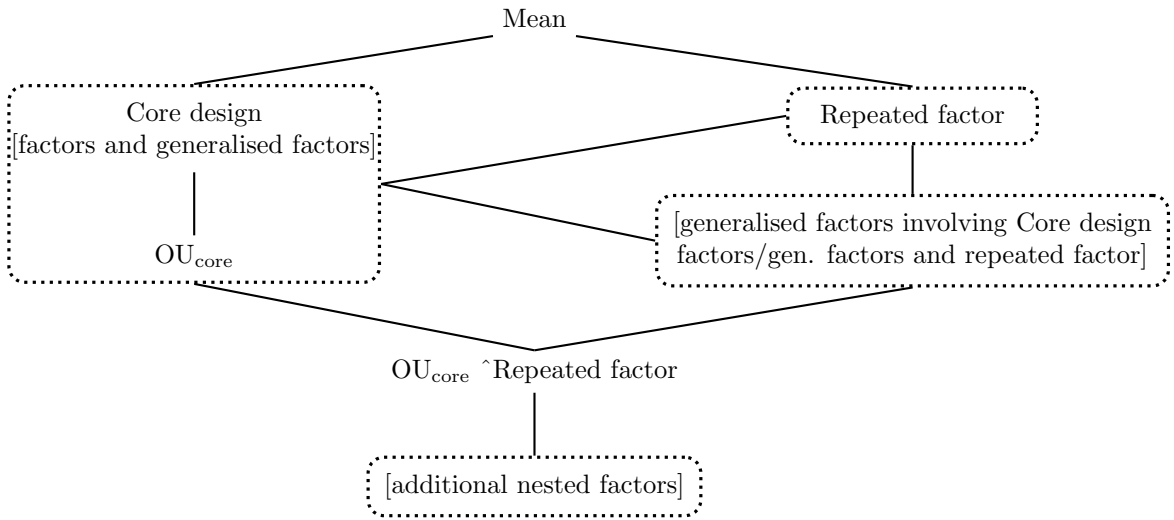
\captionof{figure}{Schematic Hasse diagram of the LS of a nested repeated measures design.}
\label{fig:nestedrepeatedmeasures}

\end{minipage}
\end{center}

\subsection*{Example 7. The Comet assay}

\noindent The Comet assay is an animal model routinely employed to assess the genotoxicity of novel pharmaceutical compounds, see \citet{bright2011}. Animals (factor: Animal) are administered either a drug treatment (at one of three doses) or control (factor: Drug) and the effect of the drug treatment on the nucleus of individual cells is quantified for different tissues (Repeated factor: Tissue). Three samples (factor: Sample) are taken per tissue with up to 50 cells assessed per sample (factor: Cell). Samples are placed on electrophoresis gels (factor: Gel), one sample per gel, such that the cells adhere to the gel. The gels are then placed in an electrophoresis bath and are subjected to an electric current, causing broken strands of DNA in each cell to migrate out of the nucleus into a comet-like tail.

\begin{figure}
\centering
\includegraphics[width=0.9\textwidth]{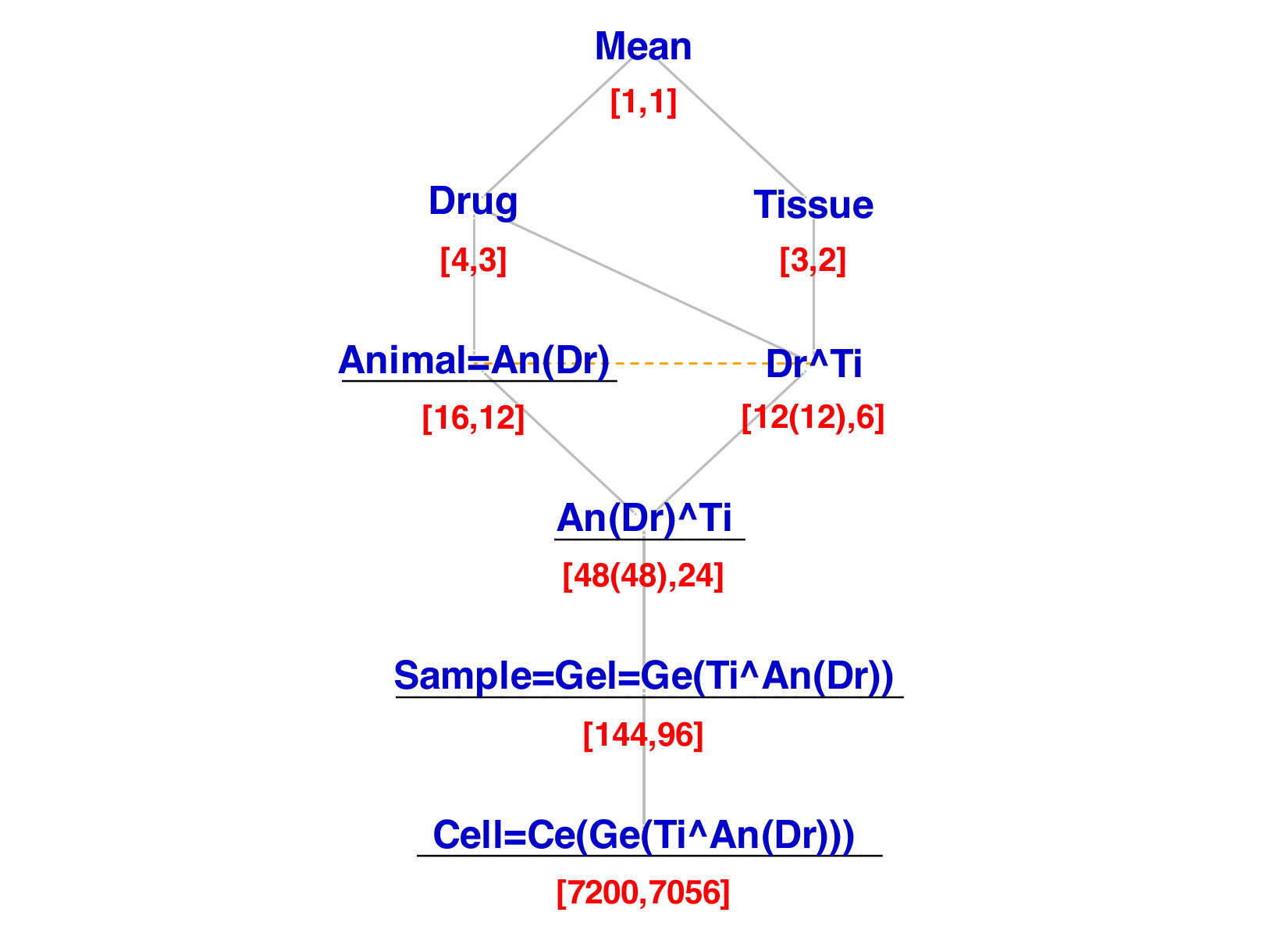}
\caption{Hasse diagram of the LS for the Comet assay repeated measures design.}
\label{fig:nestedrepeated}
\end{figure}

The design is an example of a nested repeated measure design. The Core design consists of the Animal factor nested within the Drug factor and the observational units of the Core design are the animals. As each animal is measured repeatedly, over several tissues, Animal is a Subject factor and Tissue is a Repeated factor and hence the design is a repeated measures design. The design also has an additional nested structure as there are multiple samples/gels taken for each combination of Tissue and Animal (Sample/Gel is nested within the $\text{Animal\textasciicircum Tissue}$ generalised factor) and the individual cells are nested within the samples and hence the design is defined as a nested repeated measures design. The Hasse diagram of the LS for this design is given in Figure~\ref{fig:nestedrepeated}.

\subsection{Double repeated measures design} \label{subsec:drmd}

Consider a repeated measures design, as discussed in Section \ref{subsec:rmdesign}, where the Core design is itself a repeated measures design. Such a design will involve two Repeated factors that are crossed with each other and hence is defined as a double repeated measures design. Data generated when using such a design can be analysed using a repeated measures analysis approach, but the researcher will need to model the within-subject correlation structure that results from the lack of randomisation for both Repeated factors. When fitting models for double repeated measures designs, care must be taken to specify covariance structures that reflect dependence across both repeated dimensions, while avoiding over-parameterisation, particularly in studies with limited sample sizes.

\subsection*{Example 8. Motor skill acquisition trial}

\noindent \citet{zhao2025} describe an experiment to assess the interaction between task complexity and ipsilateral primary motor cortex activation in motor skill acquisition. In the trial 48 right-handed participants (factor: Volunteer) were randomly asked to train on either a simple or complex visuomotor task (factor: Challenge). The volunteers were the assessed, pre- and post-training (factor: Test), on how well they performed a task using both their right and left hands (factor: Hand). Each volunteer is tested repeatedly as they are assessed using pre-and post-training using both hands, and both of these repeats are performed in a non-random order. Hence the experiment is an example of a double repeated measures design. 

The design consisting of Volunteer and Challenge is considered a Core design, where volunteers are nested with the challenges. If we consider the Hand factor to be a Repeated factor, then this design is an example of a repeated measures design. This Core design is itself a repeated measures design, with Test as the second Repeated factor. Hence the overall design is an example of a double repeated measures design. 

\begin{figure}
\centering
\includegraphics[width=0.9\textwidth]{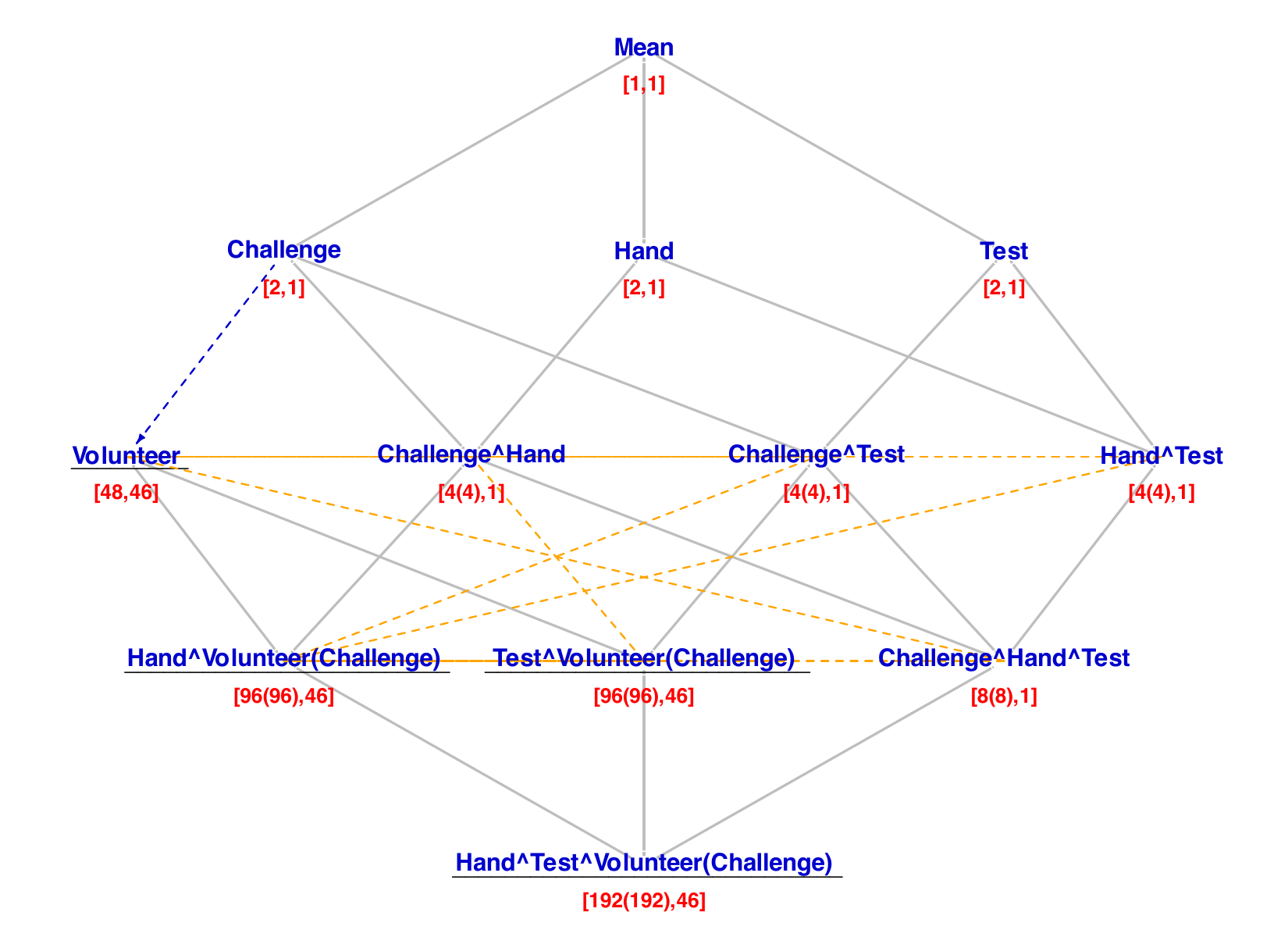}
\caption{Hasse diagram of the RLS for the Motor skill acquisition trial.}
\label{fig:doublerepeated}
\end{figure}

The Hasse diagram of the RLS is given in Figure~\ref{fig:doublerepeated}. Note that the generalised factors included in the RLS are not necessarily justified by the randomisation and hence a repeated measures mixed model will be required to fit the corresponding statistical model to the data.

\subsection{Repeated measures cross-over design}

Consider the case where a cross-over design, as described in Section \ref{subsec:crossoverdesign}, forms the Core design and additionally, within each test-period, multiple measurements are taken on each subject, see \citet[p. 246]{jones2003}. Such a design also possesses the repeated measures characteristic, with $\text{Period\textasciicircum Subject}$ as the Subject factor and the factor that indexes the repeated measurements within each test-period as the Repeated factor.

\subsection*{Example 9. Cross-over muscle recovery trial}

\begin{figure}
\centering
\includegraphics[width=0.9\textwidth]{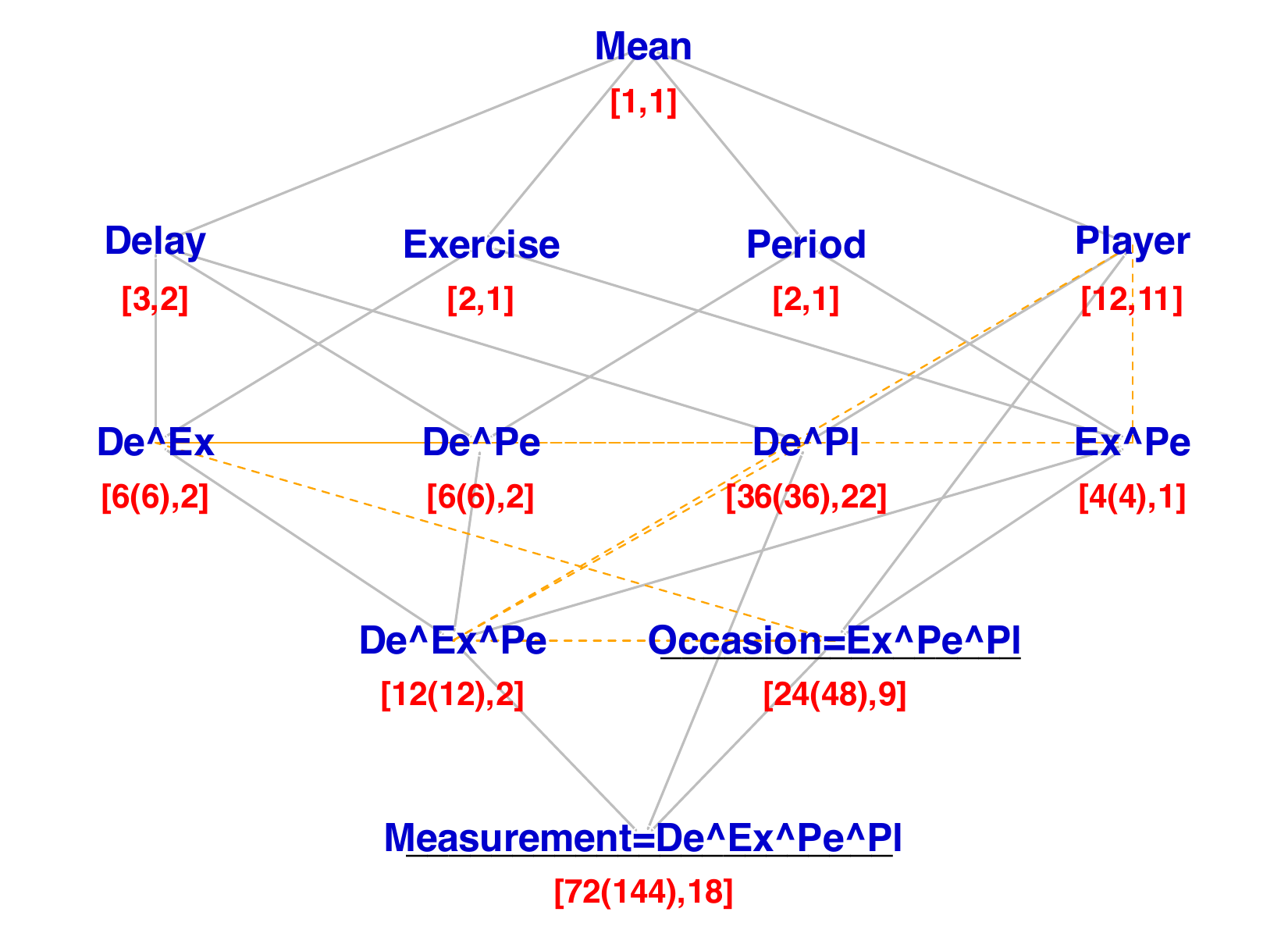}
\caption{Hasse diagram of the LS for cross-over muscle recovery trial.}
\label{fig:crossoverrepeated}
\end{figure}

\noindent A study was conducted to assess muscle damage recovery following 90 and 120 minutes of simulated soccer match-play, see \citet{field2023}. The study consisted of 12 football players (factor: Player), who completed 90 and 120 minutes of exercise (factor: Exercise) counterbalanced over two test-periods (factor: Period) in a cross-over design. The combinations of footballer and test period are defined as an occasion (factor: Occasion), a categorical factor at 24 levels. Various responses were measured (factor: Measurement) at baseline and at 24, 48 and 72 hours post-exercise (factor: Time). The design is therefore a repeated measures cross-over design, with the Exercise, Period and Player factors (all crossed with each other) defining the Core design, Occasion as the Subject factor and Time as the Repeated factor. The Hasse diagram of the LS for this study is given in Figure~\ref{fig:crossoverrepeated}.

\section{Conclusion}

The practice of using the term ‘repeated measures design’ to describe fundamentally different experiments can cause confusion to researchers that often leads to design mis-specification and hence analytical errors. To potentially reduce analytical errors, in this manuscript we are proposing the adoption of a framework for classifying experimental designs in which subjects are measured repeatedly. 

Classifying and visualising experimental designs when taking multiple measurements per subject is based on three fundamental questions: whether the subjects themselves are the experimental units, whether the experimental units are measured repeatedly, and what randomisation strategy was employed. From these questions, six design characteristics emerge. These are repeated measures, within-subject, block, split-plot, cross-over and nested. Each design characteristic is associated with structural and analytical implications. We have shown how these characteristics, individually and in combination, can be visualised using Hasse diagrams, and illustrated the framework through examples from clinical and pre-clinical applications. The value of this manuscript is that a researcher or analyst faced with a study in which subjects are measured multiple times can work through the three defining questions and arrive at a decision on the characterisation of their design. This will in turn point them to the appropriate statistical analysis, for example, a standard mixed model, a repeated measures mixed model that accounts for within-subject correlation, or something more complex when multiple characteristics are present simultaneously.

All of the models considered in this paper consist of categorical factors. In a future paper this work will be extended to include continuous factors and allow, for example, the use of random coefficient regression, ANCOVA and other polynomial regression approaches to model repeated measures data.

This work builds on the strategy described in \citet{bate2016a} and \citet{bate2016b} to encompass situations where the randomisation performed does not imply that all terms of interest should be included in the statistical model based solely on the randomisation. By placing experimental design at the centre of the modelling process, this should help the researcher not only correctly identify their experimental design but also offer them a systematic way to construct a statistical model to use in the statistical analysis that is, in some sense, justified by the design and randomisation.

\section*{Disclaimer}

The content of this article is reflective of the authors’ own personal opinions and not necessarily those of their affiliated institutions.

\section*{Acknowledgements}

The authors would like to express their sincere gratitude to Dr Marion Chatfield for her invaluable advice and encouragement throughout this research. Her guidance and constructive feedback have contributed significantly to the completion of this work.

%\section*{Supplementary material}
%
%Supplementary material associated with this article can be found in the online version at: 

\appendix
\section{Appendix A. Demonstration of the R package \texttt{hassediagrams}} \label{appendixA}
%Appendix sections are coded under \verb+\appendix+.
%
%\verb+\printcredits+ command is used after appendix sections to list 
%author credit taxonomy contribution roles tagged using \verb+\credit+ 
%in frontmatter.

This appendix illustrates how the hassediagrams R package can be used to generate Hasse diagrams (with reference to examples presented in the main text). We use Example 2 (Dog telemetry study, Section \ref{subsec:example2}) to demonstrate how to produce the Hasse diagram of the LS as illustrated in  Figure~\ref{fig:within}. Additionally, we use Example 4 (Coated implant assessment in rabbits, Section \ref{subsec:example4}) to demonstrate how to produce the Hasse diagram of the RLS as illustrated in Figure~\ref{fig:splitplot}.

The input to the package are the designs dataset stored in an Excel workbook (datasets.xlsx), where the two sheets correspond to the specific examples presented in this appendix. The sheet named "telemetry" contains the design of the Dog telemetry study and the sheet named "implant" contains the design of the Coated implant assessment in rabbits. 

Starting with the Dog telemetry study example, the dataset is read into R and passed to the \texttt{hasselayout} function, which is the function used to generate the Hasse diagram of the LS. The telemetry data frame contains one row per observation and one column per factor in the design, i.e., Dog, Treatment, Period and Measurement. The \texttt{datadesign} argument is the only mandatory argument of \texttt{hasselayout} and takes this data frame directly. The function \texttt{hasselayout} then determines the corresponding relationships between the factors (nested, fully crossed, partially crossed or equivalent) by evaluating the combinations of levels present in \texttt{datadesign}, before constructing and plotting the corresponding Hasse diagram of the LS.

By default, all factors supplied to \texttt{hasselayout} are treated as fixed. Since Measurement is a random factor in this design (and hence should be underlined in the Hasse diagram), the \texttt{randomfacsid} argument is used to identify it. The \texttt{randomfacsid} argument takes a vector of 0s and 1s, of the same length and in the same order as the columns of \texttt{datadesign}, with 1s indicating random factors and 0s indicating fixed factors. For the telemetry dataset, with columns ordered Dog, Treatment, Period and Measurement, the appropriate \texttt{randomfacsid} vector argument is \texttt{c(0,0,0,1)}, reflecting that Measurement is random, while Dog, Treatment and Period are fixed.

Finally, \texttt{larger.fontlabelmultiplier} rescales the font size used for factors displayed at levels of the Hasse diagram with few objects (four or fewer), which improves the visibility of the Hasse diagram when there are only a small number of factors, as in the case with the Dog telemetry study.

In conclusion, the Hasse diagram of the LS of the Dog telemetry study is generated using the following code:

\begin{verbatim}
library(readxl)
library(hassediagrams)

telemetry <- read_xlsx("datasets.xlsx", sheet = "telemetry")

hasselayout(datadesign = telemetry, 
            randomfacsid = c(0,0,0,1), 
            larger.fontlabelmultiplier = 2)
\end{verbatim}
and the plotted output is presented in Figure~\ref{fig:within}. 

Using the Coated implant assessment in rabbits, we demonstrate the generation of the Hasse diagram of the RLS using the \texttt{hassediagrams} package. Unlike \texttt{hasselayout}, the \texttt{hasserls} function additionally requires the randomisation performed to be specified, through the \texttt{rand.objects} and \texttt{rand.arrows} arguments. 

The "implant" excel sheet contains the four factors of this design, which are Rabbit, Time, Position and Type. Rabbit and Position are random factors and Time and Type are fixed, which is given through the \texttt{randomfacsid} argument, set here to \texttt{c(1,0,1,0)}. At first, we use the \texttt{hasselayout} function to get the structural objects.

\begin{verbatim}
library(readxl)
library(hassediagrams)

implant <- read_xlsx("datasets.xlsx", sheet= "implant")

implant_ls <- hasselayout(datadesign = implant, 
                          randomfacsid = c(1, 0, 1, 0),
                          showLS = FALSE)
implant_ls$str_objects
\end{verbatim}

Setting \texttt{showLS = FALSE} suppresses the Hasse diagram of the LS but still returns the underlying objects of the LS. The \texttt{str\_objects} element of the returned list gives the names of the structural objects, in the order required by \texttt{hasserls}. 

The \texttt{rand.objects} argument of \texttt{hasserls} is a character vector of the same length and order as \texttt{str\_objects}, giving the label to be shown on the Hasse diagram for each corresponding randomisation object, or "NULL" where there is no randomisation object corresponding to the structural object in the LS. In this example, each structural object corresponds to a randomisation object in the RLS, hence the vector is built directly from \texttt{str\_objects}, with only the labels for the fourth and sixth entries needing to be relabelled (\texttt{Rabbit} and \texttt{Position[Rabbit]}):

\begin{verbatim}
rand_spec <- implant_ls$str_objects
rand_spec[c(4,6)] <- c("Rabbit", "Position[Rabbit]")
\end{verbatim}

The randomisation performed is specified via \texttt{rand.arrows}, a two-column integer matrix in which each row defines one randomisation arrow. Specifically, the first column gives the entry number (from \texttt{rand.objects}) of the randomisation object at the start of the arrow, and the second column the entry number of the object at its end. As described in Section \ref{subsec:example4}, two randomisations were performed: rabbits were randomly assigned to the three time points ($\text{Time} \to \text{Rabbit}$), and, separately within each rabbit, the four implant types were randomly assigned to positions ($\text{Type} \to \text{Position[Rabbit]}$). Since Time, Rabbit, Type and Position[Rabbit] correspond to entries 2, 4, 3 and 6 of \texttt{rand\_spec}, these randomisations are encoded as:

\begin{verbatim}
rrandarrow <- matrix(c(2, 4,
                       3, 6),
                     ncol = 2,
                     byrow = TRUE)
\end{verbatim}
with the second column entry in each row required to be larger than the first, so that the arrows point downwards on the diagram.

The Hasse diagram of the RLS is then generated by:

\begin{verbatim}
hasserls(datadesign = implant,
         rand.objects = rand_spec,
         rand.arrows = randarrow,
         randomfacsid = c(1,0,1,0),
         larger.fontlabelmultiplier = 2)
\end{verbatim}
and the plotted output is presented in Figure~\ref{fig:splitplot}.


\bibliography{refs}

\end{document}